\documentclass[aps,prl,twocolumn,superscriptaddress,preprintnumbers,longbibliography,nofootinbib]{revtex4-1}

\usepackage{amssymb}
\usepackage{graphicx}
\usepackage{amsmath}
\usepackage{hyperref}
\usepackage{xspace}

\usepackage{color}
\definecolor{darkblue}{rgb}{0,0,0.5}
\definecolor{darkred}{rgb}{0.5,0,0}
\definecolor{darkgreen}{rgb}{0,0.5,0}

\newcommand{\be}{\begin{equation}}
\newcommand{\ee}{\end{equation}}

\allowdisplaybreaks

\begin{document}

\preprint{}

\title{Covariantizing Phase Space}

\author{Andrew J.~Larkoski}
\email{larkoski@reed.edu}

\affiliation{Physics Department, Reed College, Portland, OR 97202, USA}

\author{Tom Melia}
\email{tom.melia@ipmu.jp}
\affiliation{Kavli Institute for the Physics and Mathematics of the Universe (WPI), UTIAS, The University of Tokyo, Kashiwa, Chiba 277-8583, Japan}

\begin{abstract}
\noindent 
We covariantize calculations over the manifold of phase space, establishing Stokes' theorem for differential cross sections and providing new definitions of familiar observable properties like infrared and collinear safety. Through the introduction of explicit coordinates and a metric we show phase space is isomorphic to the product space of a simplex and a hypersphere, and we identify geometric phenomena that occur when its dimensions are large. These results have implications for fixed order subtraction schemes, machine learning in particle physics and high-multiplicity heavy ion collisions.
\end{abstract}

\maketitle

Relativistic $N$-body phase space is the manifold on which essentially all calculations in a perturbative quantum field theory take place. $S$-matrix elements are functions that live on the phase space manifold and can exhibit divergences on degenerate subspaces.  Experiments extrapolate smooth probability distributions from discrete, finite data of particles' momenta.  The first step of any Monte Carlo for fixed-order calculation or parton shower simulation involves the sampling of points on phase space.  In each of these cases, phase space itself is often treated as the background on which the calculations take place, with little focus on its intrinsic geometry.  

In this Letter, we present a covariant description of phase space and elucidate some of its novel geometric properties. 
Our aim is to demonstrate that a deeper understanding of phase space can enable the identification of restrictions on differentiable functions that can live on it, bring new interpretations of fundamental quantities like differential cross sections, and broaden the questions that can be asked of particle physics data.

We establish the application of Stokes' theorem to differential cross sections, viewing observables as providing foliations of the phase space manifold. 
This sheds new light on the criteria of infrared and collinear (IRC) safety and additivity of an observable.
These properties play a special role in massless gauge theories in four dimensions---only through the calculation of IRC safe observables do divergences from unresolved collinear or very low energy particles exactly cancel~\cite{Bloch:1937pw,Kinoshita:1962ur,Lee:1964is,Ellis:1991qj}.
Making a mathematically rigorous statement of precisely how IRC safety constrains observables is known to have problems \cite{Banfi:2004yd}.  
Much of the challenge is related to the technical fact that real and virtual divergences only need to strictly cancel in the exact soft and/or collinear limit, but a lack of a smoothness can render perturbative predictions pathological slightly away from these limits.
Nevertheless, some progress has been made by either restricting to a smaller class of observables or exploiting smoothness properties of the space of collections of particles equipped with a metric \cite{Komiske:2020qhg}.
We make a conjecture for a definition of IRC safety based on the validity of Stokes' theorem.

Establishing the phase space manifold is also important for applications of machine learning in particle physics \cite{Larkoski:2017jix,Guest:2018yhq,Radovic:2018dip,Albertsson:2018maf,Carleo:2019ptp}. 
The space in which the data input to the machine lives can be used to optimize its architecture, as exploited in convolutional or recurrent neural networks, and a neural network equivariant under the Lorentz group has recently been constructed \cite{Bogatskiy:2020tje}.  
Recent work has shown that phase space in four dimensions is a Stiefel manifold modulo the little group \cite{Henning:2019mcv,Henning:2019enq}, see also~\cite{Cox:2018wce}. Optimization and machine learning on Stiefel manifolds is well-explored in other fields, particularly for pattern recognition, e.g.~\cite{Edelman1998TheGO,doi:10.1162/089976699300015990,doi:10.1162/089976601750265036,990925,NIPS2004_2646,NISHIMORI2005106,4587733}.

We introduce explicit global coordinates that enable the construction of a metric and other quantities on phase space, providing essential input and new ways of organizing data to machine learning applications.
Another promising application of  explicit metrics on phase space is to  provide natural distance measures that can act as regularization observables; such observables are employed in techniques~\cite{Frixione:1995ms,Catani:1996vz,Weinzierl:2003fx,Anastasiou:2003gr,Kilgore:2004ty,GehrmannDeRidder:2005cm,Somogyi:2006da,Catani:2007vq,Czakon:2010td,Boughezal:2015dva,Gaunt:2015pea,Cacciari:2015jma,Caola:2017dug,Herzog:2018ily} to isolate soft and collinear divergences in modern efforts to push calculations in perturbation theory to high orders in QCD.  

We also establish geometric phenomena that occur on the phase space manifold when particle multiplicity is large, and derive new geometric test statistics in this limit. Specifically, the `curse of dimensionality'  forces the phase space volume to concentrate at the boundaries of phase space---we show  in our explicit coordinate system that this implies lightcone momenta of particles are  squeezed to the boundaries of a simplex. Such high dimensional geometry has a natural application in heavy ion physics, where multiplicities are large.

\begin{figure*}
\begin{center}
\includegraphics[width=15cm]{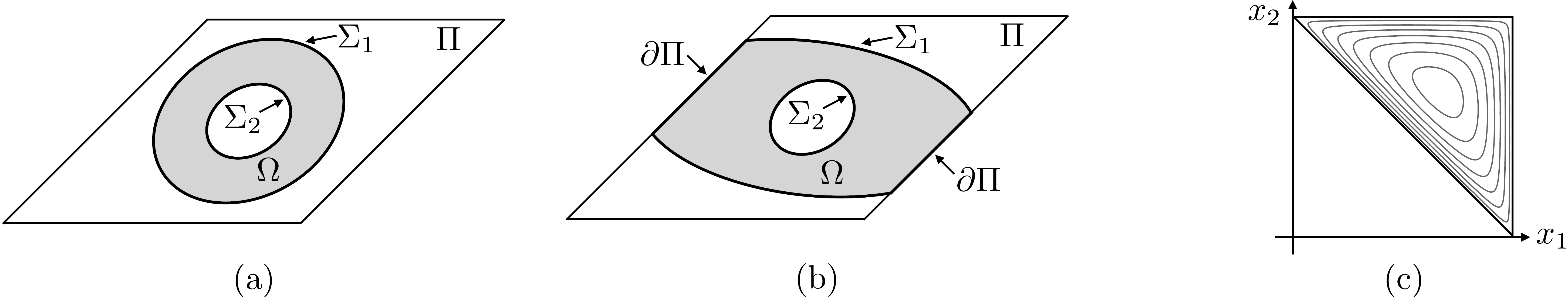}
\caption{(a) Stokes' theorem applied to a subvolume $\Omega$ of $\Pi$ bounded by two hypersurfaces $\Sigma_1$  and $\Sigma_2$ that are foliations defined by observable values ${\cal O}_1$ and ${\cal O}_2$. (b) The case for which the boundary of subvolume $\Omega$ includes part of the boundary $\partial \Pi$ of the full space.  (c) $C$-parameter foliation of three-body phase space in the $x_i$ coordinates introduced in the text.\label{fig}}
\end{center}
\end{figure*}

Let $\Pi$ denote the phase space manifold of dimension $D$.  Anticipating a covariant description, we introduce  local coordinates $x^i$ and metric $g$ on $\Pi$, and write the phase space measure in terms of the metric as  $d^D x\, \sqrt{g}$. We consider an integral of a function $M(x)$ over phase space, restricted to a hypersurface $\Sigma$, defined by $h(x)={\cal O}$, via a $\delta$-function,
\begin{equation}
F({\cal O}) = \int_\Pi d^D x\, \sqrt{g} \, M(x)\, \delta\left({\cal O} - h(x)\right)\,.
\label{eq:xsecdelta}
\end{equation}
If $M( x)$ is a squared $S$-matrix element, then $F({\cal O})$ is interpreted as a differential cross section for ${\cal O}$, as defined by the function $h(x)$ on phase space.  This notation makes the possible multiple real and virtual contributions to the matrix element implicit.
This integral form also describes observables defined by smooth weights on phase space, like the energy-energy correlation function \cite{Basham:1978bw}.  
At this point we also note that as $M(x)\ge0$, it is possible to interpret the quantity $\sqrt{g M^2}$ as a volume form, and a metric be associated with the dynamical theory; we return to this interesting possibility in the below.

The $\delta$-function has the effect of turning Eq.~\eqref{eq:xsecdelta} into an integral of a vector $V^i$  over the hypersurface $\Sigma$. To see this, we change to coordinates $(h,\tilde{x}^a)$, defining 
the induced metric on $\Sigma$,
\be
\tilde{g}_{ab} = g_{ij} \frac{\partial x^i}{\partial \tilde{x}^a}  \frac{\partial x^j}{\partial \tilde{x}^b} \,.
\ee
Using the identity $\text{det}(g_{ij}) = \text{det}(\tilde{g}_{ab})/g^{hh}$, where $g^{hh}=g^{ij}\nabla_i h \nabla_j h$,  introducing the  normal covector to the hypersurface, $N_i=(\nabla_i h)/(g^{kl} \nabla_k h \, \nabla_l h)^{1/2}$, and performing the now trivial integral over the $\delta$-function in these coordinates, it follows that Eq.~\eqref{eq:xsecdelta} becomes
\be
F({\cal O})= \int_\Sigma d^{D-1} \tilde{x}\, \sqrt{\tilde g} \, N_i  \,V^i \,,
\label{eq:xsec2}
\ee
where we define the vector
\be
V^i =    M(x) \,\frac{g^{i j} \nabla_j h(x)}{g^{kl}\nabla_k h(x) \, \nabla_l h(x) } \,.
\label{eq:vectorv}
\ee
This vector $V^i$ describes a flow in phase space along the gradient of observable ${\cal O}$:
\begin{equation}
M(x)\frac{dx^i}{d{\cal O}} = V^i\,.
\label{eq:Vflow}
\end{equation}
We will show shortly that such an interpretation is useful in classifying properties of observables.

Eq.~\eqref{eq:xsec2} is now in a form where Stokes' theorem, in covariant form, can be readily applied. Consider a volume $\Omega$ in $\Pi$ with closed boundary  $\partial \Omega$ defined by two hypersurfaces $\Sigma_1$ and $\Sigma_2$, corresponding to two values of the observable $h={\cal O}_1$ and  $h={\cal O}_2$ (see Fig.~\ref{fig}\,a), then we have, by Stokes
\begin{align}
F({\cal O}_1)-F({\cal O}_2)&=\int_{\Sigma_1} d\sigma_i  \,V^i -\int_{\Sigma_2} d\sigma_i  \,V^i 
\nonumber \\
&= \int_\Omega d^D x \sqrt{g}\,D_i\,V^i \,,
\label{eq:xsecdiv}
\end{align}
where the definition of the surface element $d\sigma_i$ can be inferred from Eq.~\eqref{eq:xsec2}, and the covariant divergence is given by
\begin{align}
D_i V^i = \nabla_iV^i + \Gamma_{ij}^j V^i\,,
\end{align}
with the contracted Christoffel symbols 
\begin{align}
\Gamma_{ij}^j = \nabla_i \log \sqrt{g}\,.
\end{align}
In general, a closed boundary on phase space could involve (subsets of) the boundary of phase space---see Fig.~\ref{fig}\,b for an illustration.

As an illustrative example, we consider 
 the foliation of  phase space by the $C$-parameter \cite{Parisi:1978eg,Donoghue:1979vi,Ellis:1980wv} in $e^+e^-\to q\bar q g$ events. Defining the variables
$
x_i=2p_i\cdot Q/Q^2\,,
$
where $p_i$ is the momentum of particle $i=1,2,3$ and $Q$ is the total momentum vector of the collision, we choose three-body phase space coordinates ($x_1$, $x_2$).  The $C$-parameter sets 
\begin{equation}
h(x)=6\frac{(1-x_1)(1-x_2)(1-x_3)}{(x_1 x_2 x_3)}\,,
\end{equation}
with $\nobreak{x_3=2-x_1-x_2}$, and we plot its contours in Fig.~\ref{fig}\,c. In these coordinates the manifold is flat, $g^{ij}=\delta^{ij}$,  the gradient is $\nabla^i = (\partial/\partial{x_1}, \partial/\partial{x_2})$, and the vector field $V^i$ in Eq.~\eqref{eq:Vflow} is straightforwardly calculated, using the squared matrix element 
\begin{equation}
M(x) = \frac{x_1^2+x_2^2}{(1-x_1)(1-x_2)}\,.
\end{equation}
With these results, one can then verify that Stokes' theorem holds for the $C$-parameter, where the difference between the differential cross section at two different values of the $C$-parameter is described by the divergence of the vector $V^i$ over that domain.  In fact, using the known value of the leading-order differential cross section for the $C$-parameter at its maximum value \cite{Catani:1997xc}, where
\begin{equation}
F\left(C = \frac{3}{4}\right) = \frac{256}{243}\pi\sqrt{3}\,,
\end{equation}
the value of the differential cross section at a general value $C$ is
\begin{equation} \label{eq:cparamstokes}
F(C) = F\left(C = \frac{3}{4}\right)-\int_\Omega d^D x \sqrt{g}\,D_i\,V^i\,.
\end{equation}
Here, $\Omega$ is the region of phase space where the $C$-parameter takes values between $C$ and $3/4$.  This formulation of the cross section with respect to endpoint values can be generalized to other observables  where the endpoint value at a given order in perturbation theory can be easily calculated or is known to vanish, for example.

The application of Stokes' theorem to cross sections differential in an observable on phase space also enables enumeration of properties of that observable that are not obvious in its original and familiar $\delta$-function form. 
In particular, for Eq.~\ref{eq:xsecdiv} to hold for a given $M(x)$ requires the function $h(x)$ on phase space to be highly restricted. 

Most acutely, if $M( x)$ is constructed from fixed-order $N$-body matrix elements,  it generically has divergences throughout $N$-body phase space as different numbers of external particles go unresolved. 
For $M( x)$ itself to be smooth on a subvolume of phase space requires embedding lower-dimensional phase space into higher-dimensional phase space and then real and virtual divergences can be canceled point by point within the larger phase space.  
The functional form of the observable ${\cal O} = h( x)$ must respect this embedding. 
In such a case, an observable is {\it infrared and collinear safe} if the quantity $F({\cal O})$ in Eq.~\ref{eq:xsec2} is calculable on all hypersurfaces defined by $h(x)$. It follows that  the lhs of Eq.~\ref{eq:xsecdiv} is calculable. 

The textbook statement of conditions for which Stokes' theorem holds is that the  manifold  $\Omega$ is smooth, and that the $D-1$ form that is integrated over the boundary $\partial\Omega$  is smooth and has compact support on $\Omega$. 
We make the conjecture that the same conditions for Stokes' theorem  to hold in  Eq.~\ref{eq:xsecdiv} are those that dictate IRC safety, namely that the $D-1$ form that is integrated over the hypersurface $\Sigma$ in Eq.~\eqref{eq:xsec2} is smooth and has compact support on all of the phase space manifold $\Pi$, in the case when $M(x)$ is constructed from fixed order matrix elements.  

In the $C$-parameter example above, this definition of IRC safety holds because Eq.~\eqref{eq:cparamstokes} holds for all values of $C$, in particular as $\partial\Omega$ approaches the boundary (and indeed when it becomes the boundary, where one should also include the contribution of a virtual matrix element).

We leave a detailed study as to whether particularly pathological IRC safe/unsafe observables exist that evade the above conjecture (and any  potential smoothness tests for IRC safety that could be performed for a given $h(x)$) to future work, and instead focus here on a geometric definition of an important subclass of IRC observables.

We say that an IRC safe observable ${\cal O} = h(\vec x)$ is {\it additive} if the trajectory from $N$- to $N+m$-body phase space with fixed total momentum by the emission of $m$ arbitrarily soft particles flows along a gradient perpendicular to the $N$-body phase space submanifold.  
Flow along the gradient exclusively in the emitted particle phase space means that the arbitrarily soft emissions can be thought of as changing the value of ${\cal O}$ on a fixed background of $N$ particles.  
This definition of an additive observable is consistent with a form established long ago \cite{Tkachov:1995kk} and generalizes a definition from Ref.~\cite{Banfi:2004yd}.  
For example, the definition of Ref.~\cite{Banfi:2004yd} can be stated in the following way.  
Let $\tau(\{p\})$ be an IRC safe observable that depends on a set of particle momenta $\{p\}$.  Then, $\tau$ is additive if there is a subset of momentum $\{\tilde p\}$ on which $\tau(\{\tilde p\}) = 0$ and $m$ additional emissions remain in the soft and collinear region which can be accomplished by scaling their momentum by a parameter $v$.  Then, in the limit that $v\to 0$, the observable takes the form
\begin{equation}
\lim_{v \to 0} \tau\left(\{\tilde p\},\kappa_1(v\zeta_1),\kappa_2(v\zeta_2),\dotsc,\kappa_m(v\zeta_m)\right) = v\sum_{i=1}^m \zeta_i\,,
\end{equation}
where $\zeta_i$ is the functional form that $\tau$ takes on particle $i$ and $\kappa_i(\zeta_i)$ is a momentum function that translates the value of $\tau$ to the realization of momentum of particle $i$.  This definition demonstrates that the hard particle momenta $\{\tilde p\}$ are completely unaffected by the $m$ soft and collinear particles, and so indeed corresponds to flow along a gradient perpendicular to the phase space manifold for momenta $\{\tilde p\}$.  This definition of additivity has the further requirement that soft particles individually contribute to the observable, while our definition just requires the value of the observable in the soft limit to exclusively be a function of the soft momenta.  Modern jet grooming algorithms, for example, can enforce correlations between the relative angle and/or energy of soft particles, while still retaining all of the nice calculability properties of additivity \cite{Frye:2016aiz}.  Because of their nice properties, additive observables are among the most widely studied and include thrust \cite{Brandt:1964sa,Farhi:1977sg}, the $C$-parameter, (recoil-free) angularities \cite{Berger:2003iw,Almeida:2008yp,Ellis:2010rwa,Larkoski:2014uqa}, $N$-(sub)jettiness \cite{Brandt:1978zm,Stewart:2010tn,Kim:2010uj,Thaler:2010tr,Thaler:2011gf}, energy correlation functions \cite{Banfi:2004yd,Larkoski:2013eya,Moult:2016cvt}, energy flow polynomials \cite{Komiske:2017aww}, among others.

As a simple example of our definition of additivity, we consider the angularities $\tau^{(\alpha)}$ measured with respect to the final state momentum or thrust axis \cite{Berger:2003iw,Almeida:2008yp,Ellis:2010rwa}, which for three-body phase space in the above coordinates, assuming the ordering $x_1, x_2\leq x_3$, can be expressed as
\begin{align}\label{eq:angdef}
\tau^{(\alpha)} &= x_1\left(
1-\frac{1-x_2}{x_1(2-x_1-x_2)}
\right)^{\alpha/2}\\
&
\hspace{2cm}
+x_2\left(
1-\frac{1-x_1}{x_2(2-x_1-x_2)}
\right)^{\alpha/2}\,,
\nonumber
\end{align}
for parameter $\alpha>0$.  In the soft limit of $x_1\to 0$, $\tau^{(\alpha)}\to 0$, demonstrating the infrared and collinear safety of the angularities.  For this class of observables, additivity means that the expansion of $\tau^{(\alpha)}$ for $x_1\to 0$ is proportional to its derivative with respect to $x_1$, in the same limit.  That is, 
\begin{equation}
\lim_{x_1\to 0} \tau^{(\alpha)} \propto x_1\left. \frac{\partial \tau^{(\alpha)}}{\partial x_1}\right|_{x_1\to 0}\,.
\end{equation}
This can only hold if the second term in the expression of Eq.~\ref{eq:angdef}, which quantifies the recoil of the harder particle 2 away from the thrust axis, is subdominant in the $x_1\to 0$ limit.
Only for $\alpha > 1$ are these angularities additive and the emission of soft particle 1 corresponds to flow perpendicular to the two-body phase space manifold.  Thrust corresponds to $\alpha = 2$ and is therefore indeed classified as an additive observable.

The value of an additive observable ${\cal O}$ is proportional to the distance along the flow defined by $V^i$ of the particles to a lower-body phase space manifold.  This property is one of the reasons that makes additive observables especially well-suited for the application of regularization of infrared divergences.  
$N$-jettiness subtraction \cite{Boughezal:2015dva,Gaunt:2015pea} is an example of an additive observable for regularization.  Additionally, it is known that some additive observables can be interpreted as a metric distance from a lower-body phase space manifold \cite{Komiske:2020qhg}.  Our covariant definition of additivity demonstrates that all additive observables enjoy this property.

Having presented a number of covariant statements about observables on phase space, we now turn to constructing an explicit coordinate system and metric for the phase space manifold. 
We begin with  a brief review of geometrical aspects of phase space that were elucidated in Refs.~\cite{Cox:2018wce,Henning:2019mcv,Henning:2019enq}.
In conventional coordinates and normalization, the volume form for four-dimensional, on-shell, massless, $N$-body phase space in the center-of-mass frame is:
\begin{align}
\hspace{-0.2cm}d\Pi_N = (2\pi)^{4-3N}\left[ \prod_{i=1}^N d^4p_i \, \delta^+(p_i^2)\right]\! \delta^{(4)}\left(Q-\sum_{i=1}^N p_i\right)\,.
\end{align}
Here $Q = (Q,0,0,0)$ represents both the total momentum four-vector and the total energy in the center-of-mass frame, and $\delta^+(p_i^2)=\delta(p_i^2)\Theta(p^0_i)$.  Our first step 
is to rescale all momenta by the center-of-mass energy: $p_i \to Qp_i$.
The on-shell $\delta$- and $\Theta$-functions can be trivially enforced by expressing a momentum $p$ as the outer product of spinors $\lambda^a$ and $\tilde\lambda^{\dot a}$, where
\begin{align}\label{eq:spinmommap}
(p\cdot \sigma)^{a\dot a}&=\left(
\begin{array}{cc}
p_0 - p_3 & -p_1+ip_2\\
-p_1-ip_2 & p_0 + p_3
\end{array}
\right)^{a\dot a}\\
& = \lambda^a\tilde\lambda^{\dot a} = \left(
\begin{array}{cc}
\lambda^1\tilde\lambda^{\dot 1} & \lambda^1\tilde\lambda^{\dot 2} \\
\lambda^2\tilde\lambda^{\dot 1} & \lambda^2\tilde\lambda^{\dot 2} 
\end{array}
\right)^{a\dot a}\,.\nonumber
\end{align}
Reality of momentum $p$ requires that $\lambda^a$ and $\tilde \lambda^{\dot a}$ are complex conjugates: $\tilde \lambda^* = \lambda$.

In these spinor coordinates, the on-shell integration measure for momentum $p$ becomes
\begin{equation}
d^4p \, \delta(p^2) \,\Theta(p_0) = \frac{d^2\lambda^1\, d^2\lambda^2}{{U(1)}}\,,
\end{equation}
where the division by $U(1)$ represents implicit restriction to one element of the little group action on the spinors.  The momentum conserving $\delta$-functions can be expressed most simply through construction of two $N$-dimensional complex vectors
\begin{align}
&\vec u = \left(
\lambda_1^1\,\,\, \lambda_2^1\,\,\,\cdots\,\,\,\lambda_N^1
\right)\,,&\vec v = \left(
\lambda_1^2\,\,\, \lambda_2^2\,\,\,\cdots\,\,\,\lambda_N^2
\right)\,,
\end{align}
where $\lambda_i^a$ is the $a$th ($a=1,2$) component of the spinor for the $i$th particle.  In terms of $\vec u$ and $\vec v$, the phase space volume element is compactly
\begin{align}\label{eq:ps_simp}
d\Pi_N &=(2\pi)^{4-3N}Q^{2N-4}  \frac{d^N u\,d^Nv}{{U(1)}^N}\\
&
\hspace{2cm} \times\delta\left(1-|\vec u|^2\right)\,\delta\left(1-|\vec v|^2\right)\,\delta^{(2)}\left(\vec u^\dagger \vec v\right)\nonumber\,.
\end{align}

The phase space measure describes two orthonormal $N$-dimensional complex vectors $\vec u$ and $\vec v$.  The orthonormal constraints are invariant under the action of $U(N)$ and further $U(N-2)$ acts on a subspace without affecting $\vec u$ or $\vec v$.  Thus the phase space manifold, which we denote by $\Pi_N$, can be expressed as the quotient space
\begin{align}
\Pi_N &\cong \frac{1}{{U}(1)^N}\frac{{U}(N)}{{U}(N-2)} \\
&= \frac{1}{{U}(1)^N}\frac{{U}(N)}{{U}(N-1)}\frac{{U}(N-1)}{{U}(N-2)} \nonumber\\
&=  \frac{1}{{U}(1)^N} S^{2N-1}\times S^{2N-3}\nonumber\,.
\end{align}
The quotient space $U(N)/U(N-2)$
  is the Stiefel manifold of complex two-frames in $\mathbb{C}^N$:
\begin{equation}
V_2(\mathbb{C}^N) = \frac{{U}(N)}{{U}(N-2)}\,.
\end{equation}
The quotient space $U(N)/U(N-1) = S^{2N-1}$, the $2N-1$ sphere, so the phase space manifold has topology of a product of spheres modulo the action of the little group.  

Our development of these ideas from hereon is two-fold. First, we establish the topology of phase space when the little group redundancy is eliminated. This is important for applications of machine learning, so as to remove the need for learning of redundant directions on the manifold. Second, we provide explicit global coordinates and construct a metric on phase space.

The little group action can be explicitly accounted for by `gauge fixing' (see also~\cite{Cox:2018wce} where this was done in conjunction with fixing the Lorentz frame, whereupon phase space has topology of a Grassmann manifold).  
We focus on the action of the little group on vector $\vec u$ for which its $i$th entry is
\begin{equation}
u_i=\lambda_i^1 = r_i e^{i\phi_i}\,.
\end{equation}
We can then express the integration measure for $\vec u$ mod the little group as
\begin{align}\label{eq:simplexderiv}
\hspace{-0.23cm}\frac{d^N u}{{U(1)}^N}\delta\left(1-|\vec u|^2\right)&= \frac{\prod_{i=1}^N r_i\, dr_i\, d\phi_i}{{U(1)}^N}\, \delta\left(
1-\sum_{i=1}^N r_i^2
\right)\nonumber\\
&
\hspace{-1cm}=\int \prod_{i=1}^N \left[\frac{dr_i^2}{2}\, d\phi_i\, \delta(\phi_i)\right]\,\delta\left(
1-\sum_{i=1}^N r_i^2
\right)\nonumber\\
&
\hspace{-1cm}=\frac{1}{2^N}\prod_{i=1}^N \left[d\rho_i\right]\, \delta\left(
1-\sum_{i=1}^N \rho_i
\right)\,.
\end{align}
In the first equation, we express the integration measure in polar coordinates for the components of $\vec u$.  In the second line, we use the little group invariance to explicitly fix the phases $\phi_i = 0$, and then on the third line we make the change of variables $r_i^2 = \rho_i$.  The variable $\rho_i$ is just a lightcone component of momentum, from the mapping in Eq.~\ref{eq:spinmommap}:
\begin{equation}
\rho_i = r_i^2 = (\lambda_i^1)^2 = p_{0,i}-p_{3,i} \equiv p_i^+\,.
\label{eq:lightconep}
\end{equation}
The manifold that remains after explicitly accounting for the little group is the $N-1$ simplex $\Delta_{N-1}$ 
with unit base, which represents the conservation of $+$-component lightcone momentum.  We can then express the integration measure for $\vec u$ as
\begin{equation}
\frac{d^N u}{{U(1)}^N}\delta\left(1-|\vec u|^2\right) = \frac{1}{2^N}\,d\Delta_{N-1}\,,
\end{equation}
where $d\Delta_{N-1}$ represents the flat measure on the simplex.

The other factor in the phase space measure that depends on the vector $\vec v$ can be manifestly expressed as the measure for the sphere $S^{2N-3}$.  We just outline the procedure here.  First, the $\delta$-function that enforces orthogonality of $\vec u$ and $\vec v$ can be used to eliminate the component $v_N$ so that
\begin{equation}
d^Nv\, \delta\left(1-|\vec v|^2\right)\,\delta^{(2)}\left(\vec u^\dagger \vec v\right) = \frac{d^{N-1}v}{\rho_N}\, \delta\left(1-|\vec v|^2 - |v_N|^2\right)\,,
\end{equation}
where the factor of $1/\rho_N$ is the resulting Jacobian and we have left $v_N$ component implicit in the remaining $\delta$-function.  In general, $v_N$ now has dependence on all $N-1$ other components of $\vec v$ as well as all coordinates of the simplex, so the measure is not manifestly that of the sphere.  We can transform it into the desired form by changing variables from $\vec v$ to $\vec v'$, under which the real and imaginary parts of $\vec v$ mix only amongst themselves via a real, symmetric matrix.  Conservation of the $\rho$ coordinates of the simplex ensures that this transformation has the Jacobian $J = \rho_N$, rendering the measure exactly that of the sphere.  
That is, the measure for the $\vec v$ coordinates can be expressed as
\begin{align}
&d^Nv\, \delta\left(1-|\vec v|^2\right)\,\delta^{(2)}\left(\vec u^\dagger \vec v\right) \\
&
\hspace{2cm}= d^{N-1}v'\, \delta\left(1-|\vec v'|^2\right)\equiv dS^{2N-3}\,,\nonumber
\end{align}
the measure of the $2N-3$ sphere.  Explicit coordinates for $\vec v'$ that ensure normalization $ |\vec v'|^2=1$ are
\begin{align}\label{eq:hopfcoords}
v_1' &=  e^{-i\xi_1}\cos\eta_1\,,\\
v_2' &= e^{-i\xi_2}\sin\eta_1\cos\eta_2\,,\nonumber\\
&\hspace{0.2cm} \vdots \nonumber\\
v_{N-2}' & = e^{-i\xi_{N-2}}\sin\eta_1\cdots \sin\eta_{N-3}\cos\eta_{N-2}\,,\nonumber\\
v_{N-1}' & = e^{-i\xi_{N-1}}\sin\eta_1\cdots \sin\eta_{N-3}\sin\eta_{N-2}\,.\nonumber
\end{align}
These generalize coordinates for the Hopf fibration to the embedding of $S^{2N-3}$ in $\mathbb{C}^{N-1}$.  The parameters have ranges $\xi_i\in[0,2\pi]$, and  $\eta_i\in[0,\pi/2]$, and the  volume form in these coordinates is
\begin{align}
&dS^{2N-3}=\\
&
\hspace{0.5cm}\left(\prod_{k=1}^{N-2} \cos \eta_k \sin^{2k+1} \eta_k\right) d\xi_1\cdots d\xi_{N-1}\,d\eta_1\cdots d\eta_{N-2} \,.\nonumber
\end{align}
The phase space manifold is the product of the simplex and the sphere:
\begin{equation}
\Pi_N \cong \Delta_{N-1}\times S^{2N-3}\,.
\label{eq:newtopology}
\end{equation}
The dimension of phase space is reconstructed as the sum of the simplex and sphere dimensions, $(N-1)+(2N-3)=3N-4$. Similarly the phase space volume can be especially easily derived in this framework, using well-known formulas $\nobreak{\text{Vol}(\Delta_{N-1}) = 1/{(N-1)!}}$ and $\nobreak{\text{Vol}(S^{2N-3}) = {2\pi^{N-1}}/{(N-2)!}}$.

We can further construct the line element (metric) on the phase space manifold.  As phase space is a product manifold, its line element can be constructed from the individual line elements of the simplex and sphere, requiring the resulting line element to be positive definite and produce the correct volume form for phase space.  The line element of the simplex is just the Euclidean metric in the $\rho$ coordinates:
\begin{equation}
ds^2_\Delta = \sum_{i=1}^{N-1}d\rho_i^2\,.
\end{equation}
In the Hopf-like coordinates, the line element of the sphere satisfies a recursive relationship:
\begin{equation}
ds_{S^{2N-3}}^2 = d\eta_1^2 + \cos^2\eta_1\,d\xi_1^2 + \sin^2\eta_1\,ds_{S^{2N-5}}^2\,,
\end{equation}
and the line element on $S^1$ is flat, $ds_{S^1}^2 = d\xi_1^2$.
It is trivial to extend this for systems in which energy conservation is not assumed, but note that we can always work in the frame in which the net three-momentum is zero.  A metric on unordered and arbitrary collections of particle momenta has been proposed \cite{Komiske:2019fks,Komiske:2020qhg}, but to our knowledge, this is the first that is directly constructed from the phase space manifold.

We end by presenting some observations about the geometry of phase space we derived in Eq.~\eqref{eq:newtopology} in the case where the dimension $3N-4$ is large. One of the phenomena associated with the `curse of dimensionality' is that the volume of a manifold with boundary becomes increasingly concentrated at its boundary. For the $n$-ball, for example, when $n=1000$, more than $99.99\%$ of its volume lies within $1\%$ of the surface. When little group redundancy is removed, phase space has a boundary, and in the coordinates we introduced, this is the boundary of the simplex of lightcone momenta $p_i^+$, see Eq.~\eqref{eq:lightconep}. On the boundary, one or more of the $p_i^+$ are zero. As particle multiplicity grows, so does the dimension of the simplex, and the concentration of phase space at the boundary implies that a number of particles will have close to zero $p^+_i$.

This can be made more quantitative, and general statements about how lightcone momenta are distributed around zero at large particle multiplicity can be derived purely from geometrical features of the high dimensional phase space manifold.
We assume a flat matrix element on phase space and that the number of particles $N$ is large, $N \gg 1$.
The probability of $m$ particles with $p^+$ less than $\rho_{\min} Q$ for $\rho_{\min}\ll 1/N$ is
\begin{equation}
p_m = {N \choose m}N^m\rho_{\min}^m(1-N \rho_{\min})^{N-m}\,.
\label{eq:bindist}
\end{equation}
As a binomial distribution, its mean is $\mu=N^2\rho_{\min}$ and variance is $\sigma^2 = N^2\rho_{\min}(1-N\rho_{\min})$.  This novel `large $N$' limit is still consistent with $\sigma \ll \mu$ in which a significant number of particles have very small lightcone momenta.

The assumption of a flat matrix element on phase space is motivated by strongly-coupled systems like heavy ions for which the matrix elements are expected to be smooth, non-singular distributions on phase space. The large event multiplicities suggests that the probabilities $p_m$ given in Eq.~\eqref{eq:bindist} could make interesting test statistics.  

Relaxing the assumption of a flat matrix element to one that is slowly varying means that a harmonic expansion on the Stiefel manifold/phase space (see~\cite{Henning:2019mcv,Henning:2019enq})  would quickly converge, and a small number of coefficients of that expansion would quantify interesting correlations at different angular and energy scales, as would deviations from the $p_m$ given in Eq.~\eqref{eq:bindist}.  This scenario also suggests itself as one well-suited to the  interesting possibility we mentioned under Eq.~\eqref{eq:xsecdelta}---of incorporating the slowly varying matrix element $M(x)$ itself into the metric. The resulting geometry would encapsulate both phase space and dynamics, and it would be fascinating to study geodesics and the observable flows per Eq.~\eqref{eq:Vflow} in this space. This would be the ultimate promotion of phase space---from a background upon which calculations take place to being geometrically entwined  with a theory's dynamics.

\begin{acknowledgments}
A.L.~thanks the IPMU for support and hospitality where this work was initiated.  
We thank Ben Nachman for emphasizing the importance of the manifold of input data in machine learning, Patrick Komiske, Eric Metodiev and Jesse Thaler for comments on the manuscript and discussions regarding the relationship to the metric of Ref.~\cite{Komiske:2020qhg},
and Peter Cox, Ian Moult, Duff Neill, and Mihoko Nojiri for useful discussions and comments. 
T.M.~is supported by the World Premier International Research Center Initiative (WPI) MEXT, Japan, and by JSPS KAKENHI grants JP18K13533, JP19H05810, JP20H01896 and JP20H00153.
\end{acknowledgments}

\bibliography{ps_manifold}

\end{document}